# Multiview microscopy of single cells through microstructure-based indirect optical manipulation

*Gaszton Vizsnyiczai[1,†], András Búzás[1,†], Badri Lakshmanrao Aekbote[1,2], Tamás Fekete[1], István Grexa[1], Pál Ormos[1], Lóránd Kelemen[1,*]*

[1]Institute of Biophysics, Biological Research Centre, Hungarian Academy of Sciences, Temesvári krt. 62, Szeged, 6726, Hungary

[2]School of Engineering, James Watt South Building, University of Glasgow, Glasgow, G12 8QQ, UK

\* corresponding author, tel: 36-62-599-600 x419, email: kelemen.lorand@brc.mta.hu
† these authors contributed equally

| author's name | author's email address |
|---|---|
| Gaszton Vizsnyiczai | gaszton@brc.hu |
| András Búzás | andrewbuzas81@gmail.com |
| Badri Lakshmanrao Aekbote | Badri.Aekbote@glasgow.ac.uk |
| Tamás Fekete | fekete.tamas@brc.mta.hu |
| István Grexa | grexai926@gmail.com |
| Pál Ormos | ormos.pal@brc.mta.hu |
| Lóránd Kelemen | kelemen.lorand@brc.mta.hu |

**Running title:** Microscopy with indirect cell manipulation

Vizsnyiczai et al.: Microscopy with indirect cell manipulation**Abstract**

Fluorescent observation of cells generally suffers from the limited axial resolution due to the elongated point spread function of the microscope optics. Consequently, three-dimensional imaging results in axial resolution being several times worse than the transversal. The optical solutions to this problem usually require complicated optics and extreme spatial stability. A straightforward way to eliminate anisotropic resolution is to fuse images recorded from multiple viewing directions achieved mostly by the mechanical rotation of the entire sample. In the presented approach, multiview imaging of single cells is implemented by rotating them around an axis perpendicular to the optical axis by means of holographic optical tweezers. For this, the cells are indirectly trapped and manipulated with special microtools made with two-photon polymerization. The cell is firmly attached to the microtool and is precisely manipulated with 6 degrees of freedom. The total control over the cells position allows for its multiview fluorescence imaging from arbitrarily selected directions. The image stacks obtained this way are combined into one 3D image array with a special image processing algorithm resulting in isotropic optical resolution. The presented tool and manipulation scheme can be readily applied in various microscope platforms.

2

1. Introduction

Optical trapping has developed into a widely used approach in manipulation of biological objects. The possibility of handling microscopic particles without mechanical contact offers advantages in practically every area of experimental biology. The typical fundamental parameters of optical traps - micrometer trap size and pN exerted force - make it ideally suited to manipulate biological objects in 3D as well as to measure forces exerted by biological systems. In fact, thanks to continuous development the state of the art represents displacement measurements with sub-nanometer accuracy[1] and forces with femtoN sensitivity[2]. Among countless application examples, like study of DNA and DNA-associated proteins[3,4], mechanical protein folding-unfolding[5,6], study of molecular motors at the single-molecule level[7,8], etc., optical trapping offers great advantages in the manipulation of whole cells, too. Optical trapping of whole cells has been introduced in the very early phase of the development of the approach[9], and has been pursued subsequently[10,11]. However, it soon became apparent that direct optical trapping of live cells suffers from serious issues. Cells are typically characterized by a low refractive index contrast to water, which results in low optical trapping forces. The structural complexity of cells results in optical inhomogeneity that makes optical manipulation a complicated procedure[9,12]. Trapping occurs at high refraction index organelles: the actual point of fixation cannot be predicted. In conclusion, in the case of direct optical manipulation of cells the trapping strength and position are not known and cannot be precisely controlled. Furthermore, the high laser intensity at the focus is potentially harmful to the cell[13-17]. While careful selection of the wavelength of the trapping light can reduce the damage, cell viability is always a problem and it has to be assessed in every experiment.

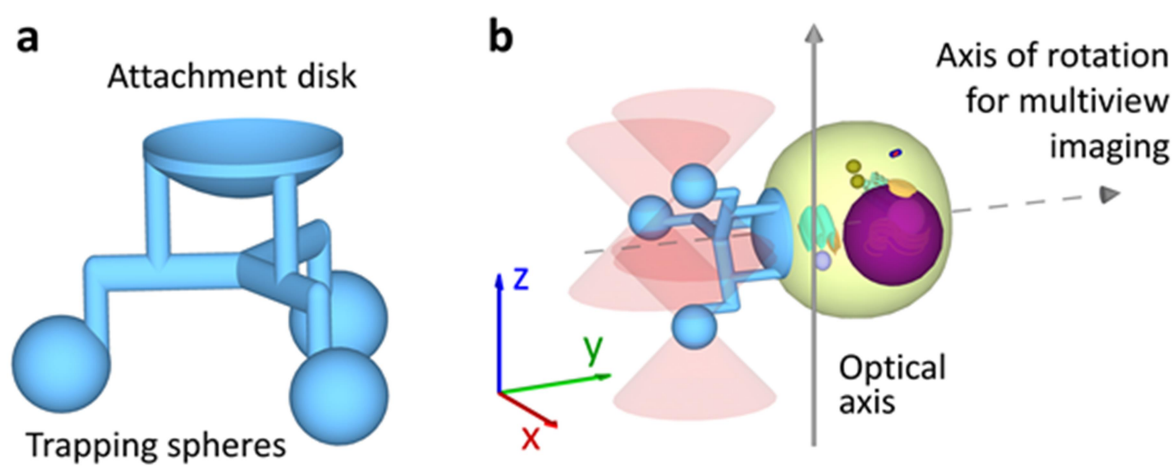

**Figure 1. The scheme of the polymer micromanipulator and its application to indirectly manipulate a single cell for multiview microscopy. a** The model of the manipulator showing its main functional parts. **b** The spatial arrangement of the manipulator-cell couple in the sample space relative to the optical axis of the trapping and observing objective. Pink cones indicate the trapping beams. The cell is rotated around the dashed-line axis for the multiview microscopyc obsrevations (parallel to axis y).

The above mentioned inherent problems of direct cell trapping can be eliminated by applying indirect manipulation, that is, by decoupling the trapping light from the live cell to be manipulated. In this scenario, an intermediate object is attached to the cell, and the trapping light interacts with this object. So far this was only achieved by the application of silica microbeads attached to the cell, as demonstrated for instance in studies to investigate mechanical properties of red blood cells[18]. Even larger improvement can be achieved with the use of purpose-built manipulators as intermediate objects (for an example, see Fig. 1a). Such microtools can be fabricated with an optimized shape for high precision trapping, where a set of small radius spherical handles provide well-defined trapping points and large trapping forces by using high refractive index materials[19]. Photodamage can be prevented for harmless cell manipulation by attaching the cell to a structural element that is positioned micrometers away from the trapping beams. Recently we introduced such an indirect optical micromanipulation method[20,21] where single cells could be manipulated with 6 degrees of freedom (6DoF) by the use of shape-optimized microtools produced by two-photon polymerization (TPP). The microtool is operated by holographic optical tweezers (HOT, Fig. 1b) and the cell is attached to them by biochemical means. We have also shown that positional accuracy and stability in the range of sub-100 nm can be routinely achieved. In this study, we present an application of the approach that highlights the benefits of the indirect manipulation method using shape-optimized microtools. To demonstrate its capabilities, we apply the method to improve the imaging of a wide-field fluorescent optical microscope by substantially increasing the axial resolution.

The elongated point spread function of an optical microscope results in a limited axial resolution of the three dimensional image, the transversal resolution being typically 3-5 times better than that in the direction along the optical axis. This disadvantage originates in the fundamental properties of optical imaging, but new procedures are being elaborated continuously to reduce this most unwelcome effect. A characteristic and very effective method is 4Pi microscopy[22], where the sample is observed by two large numerical aperture objectives positioned confocally in opposite direction both for delivering fluorescence excitation and emission detection. However, this arrangement is too complex for a number of applications, and different modifications of the classic single-sided, one objective microscopy also yields reasonable alternative solutions. Examples are those of structured illumination[23], the application of special pupils[24], spinning disk[25], or just smart combination of the refractive indices of the surrounding media[26]. A straightforward way to eliminate the anisotropic resolution is the reconstruction of the image acquired by observations from different directions, either using tomography[27,28], or superposition of 3D image stacks[29].

The latter procedure, often termed as multiview microscopy where image stacks obtained from different observation directions are used[30], depends on an effective method to access these directions of the sample. Such methods have already been applied successfully for multicellular systems where the sample was embedded in a gel matrix and rotated mechanically around an axis perpendicular to the optical axis[31,32], or a complex objective system is built around the sample[33]. When imaging single cells, an appropriate manipulation scheme has also to be applied that makes the needed orientations possible. The versatile technique of optical micromanipulation has already been used for rotational manipulation at the single cell level. This method naturally works in an aqueous environment, where the cells can be easily studied under physiological conditions. The torque for such rotation may originate in the angular momentum of the light forming the trap[34,35], the shape of the trapping light beam[36,37] or the shape of the trapped object[38,39]. 3D imaging of single cells was already approached with direct cell trapping by holographic optical tweezers, built on a spinning disk confocal microscope when several traps grabbed optically dense spots inside a cell enabling its positional manipulation[29]. Recently, an optofluidic cell rotator was also introduced for improved cell imaging[40], where the cell is held by counter propagating optical tweezers in a microfluidic channel such that it is free to rotate around an axis perpendicular to the channel direction and the fluid flow. Since this axis was not in the center of the channel, the non-symmetric force due to the parabolic flow velocity profile caused continuous, uninterrupted rotation of the cell during observation.

However, for the true gain in the size of the point spread function and in achieving isotropic resolution for 3D single cell imaging, a more precise and better controlled manipulation scheme with appropriate accuracy has to be applied. In this work, we show that our indirect optical manipulation method that is based on a purpose-built polymer microtool (Fig. 1a) is capable of positioning cells with the needed precision and stability. Applying this manipulation technique, we can record wide field z-scan image stacks taken from different viewing directions (rotation axis shown on Fig. 1b). Fusing these recordings into a single 3D stack, we realize fluorescence imaging with substantially improved axial resolution, which we validated with cell-attached fluorescent nanobeads. We show that our method can reconstruct the 3D structure of fixed white blood cells (K562) with fluorescently stained mitochondria. The resultant images of the improved resolution demonstrate the power of the indirect cell manipulation method.

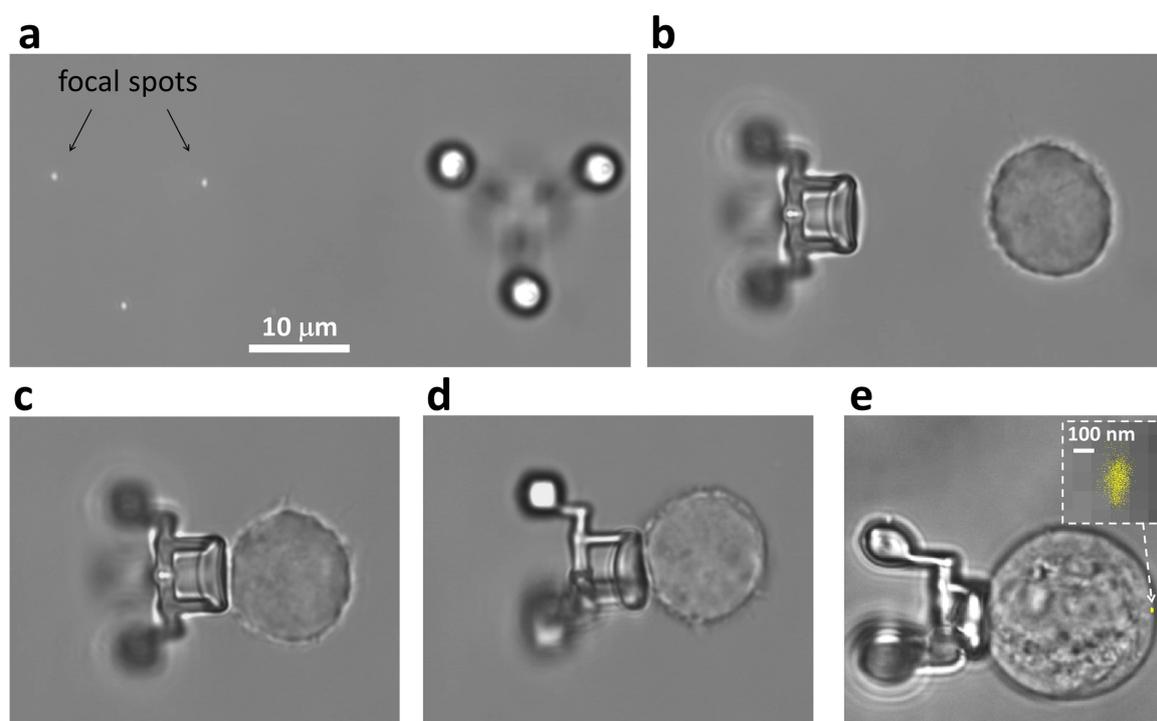

**Figure 2. The process of cell attachment to a microtool and their total positional control. a** Bright-field image of the three trapping foci (three bright spots on the left side) and a yet untrapped microtool (on the right side); the attachment disk and connector rods are defocused. **b** The microtool trapped and oriented with its disk towards the cell that sits on the bottom of the microfluidic channel. **c** The cell is attached to the microtool and elevated from the supporting glass. **d** The indirectly trapped cell is rotated by 90 degrees relative to its orientation on panel **c**. **e** The positional distribution of a given point on a fluctuating trapped cell; the insert shows the distribution zoomed-in.

2. **Results**

*Indirect optical manipulation with 6 degrees of freedom*

The most important result of the present work is to demonstrate that the 6 degrees of freedom manipulation of single cells using polymer microstructures and HOT can easily support multiview microscopy to increase optical resolution. The cells are attached to the microtools within seconds (Fig. 2a-c) and the subsequent manipulation allows for holding them firmly, translating them within the field of view in any direction at tens of micrometers distance, and most importantly in this case, allows for their precise and total rotational control. The cells can be rotated around any of the three axes quickly so that the full solid angle range is accessed within seconds, therefore the cell can be easily viewed from any direction. The indirect manipulation also ensures that the cell exposure to the trapping beams that grab the manipulators by their spherical parts becomes negligible when the cell-structure complex is aligned perpendicularly to the optical axis; in other orientations the cell is still micrometers away from the intense foci of the beams. Although the manipulator's attachment disk touches the cell directly, it covers only a small, 5.5% surface area considering a typical cell diameter of 16 μm; this ratio can be further reduced if the contact area for live cells is an issue. Fig. 2c-e demonstrates that the trapping foci are about 10 micrometers away from the cell at all orientations for this particular structure.

Even though the trapped structure-cell complex is firmly held, it of course undergoes thermal fluctuation. This motion in our case may blur the observed image of the cell therefore it needs to be accounted for. The positional stability of the cells was measured in order to evaluate the degree of this blur during an image acquisition session (Fig. 2e). The cells were held in place by a steady trap and 2000 frames were recorded with 1 ms exposition time. Analyzing the cell positions we found that the half width of the position fluctuation of the part of the cell that is the remotest from the manipulator and fluctuates the most is around 80-90 nm along the symmetry axis and 140-160 nm in the perpendicular direction; from these values the angular fluctuation of the structure-cell couple was calculated to be 0.34 degrees on average. This fluctuation poses an upper limit for the degree of image blur of any observable fluorescent point source in the cell.

*Resolution improvement along the optical axis*

For multiview imaging, the firmly trapped cell-microtool couple is first oriented with its symmetry axis perpendicular to the optical axis. An image stack is then recorded by translating the cell along the optical axis. Multiview recordings are realized by repeating this scan with different rotational orientations around the symmetry axis. For further details, see Methods.

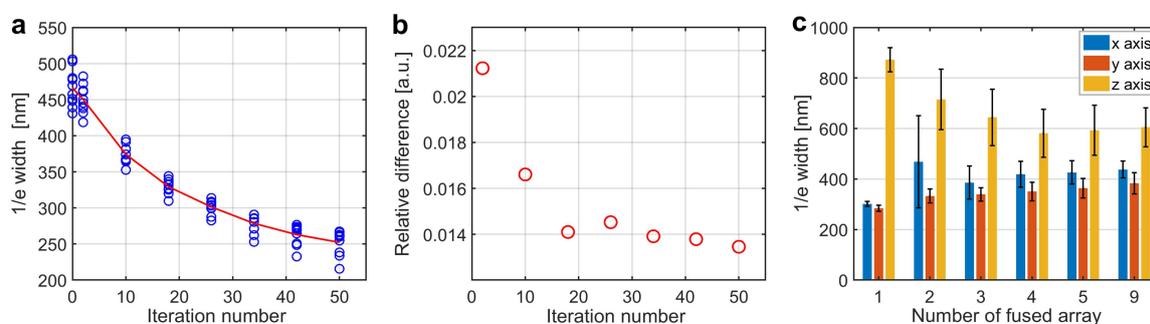

**Figure 3. Evaluation of the required deconvolution iteration number and of the required number of 3D stacks in the fusion step. a** The width of eight selected bead images along the x axis as the function of decovolution iteration number. The widths were determined with a Gaussian fit to an intensity trace taken across the maximum intensity pixels of the bead images; open circles: calculated widths; the red line connects the average of the widths. **b** Relative differences between successive arrays, calculated with eq. 1, as the function of decovolution iteration number. **c** Average widths of six selected bead images along the 3 coordinate axes after fusion of various number of aligned stacks (1 means a single deconvolved image). Fourier-based fusion was used and the width was determined with a Gaussian fit similarly as in panel **a**.

The enhancement of the axial resolution was checked with cells decorated with 100 nm diameter fluorescent beads. Our goal was to obtain images of these beads that have uniform sizes along all three dimensions. The beads attached to the outside of the cell membrane served as point-like sources characterizing the resolution improvement, a standard procedure in multiview microscopy[33]. Nine image sequences were recorded at orientations of multiplies of 22.5 degrees, covering the range from 0 to 180 degrees. First, these nine 3D data arrays were deconvolved with the measured point spread function (PSF) of our system using the Richardson-Lucy (RL) algorithm. The halt of this iterative algorithm for the bead images was determined by *i)* comparing the average width of the deconvolved bead images along the x and y axes obtained by a Gaussian fit to the lateral resolution of our system ($w = 0.61 \cdot \lambda / NA = 310$ nm, where $\lambda = 610$ nm and $NA = 1.2$) and by *ii)* calculating the relative difference between the i$^{th}$ and the (i-1)$^{th}$ deconvolved image according to the following formula:

$$\sum_x^N \|St_i(x) - St_{i-1}(x)\| / \sum_x^N St_{i-1}(x), \qquad \text{eq. 1.}$$

where $St_i(x)$ is the intensity of the image stack at the pixel of coordinate *x* after the i$^{th}$ iteration and *N* is the number of pixels. The halt of the iteration was linked to that number where the average 1/e width of the Gaussian fits of the bead images reaches 310 nm (Fig. 3a) and where the relative difference (Fig. 3b) levels; therefore we stopped the deconvolution after 26 iterations. Fig. 4a and b shows the view of a 3D image stack from a direction perpendicular to the optical axis (i.e. axis x on Fig. 1b) after recording and after deconvolution. It is apparent that deconvolution removes most of the out-of-focus intensities (noise) but keeps the elongated, most intense parts of the bead images.

After deconvolution, the image stacks were spatially registered resulting in precisely overlapping bead images, which is a pre-requisite for the successful fusion. The center positions of the bead images on each aligned stack differ slightly more than one pixel from the respective image centers on the reference stack, showing the effectiveness of the registration procedure. We reasoned the number of directions to fuse according to Fig. 3c that shows the widths (at 1/e of the maximum) of selected fused bead images as determined with a Gaussian fit. The fusion of only 2 (perpendicular) directions already reduces the widths along the optical axis from about 880 nm to about 708 nm. Increasing the number of fused images results in a decrease of the axial width to about 600 nm for 9 orientations, displaying a weak minimum (580 nm) at 4 fused directions (0°, 45°, 90° and 135°). Interestingly, the lateral widths slightly increase with the number of fused arrays but level after 4 arrays. Although there was no significant difference between 4, 5 or 9 directions, because of the presence of the minimum of the axial width we used the above 4 directions to assess the resolution improvement.

Considering fusion, there's no standard theory for the fusion of the image stacks of the various directions, several methods are used in the literature. We chose to use two methods to illustrate their effect on resolution improvement: the arithmetic averaging of the stacks in the real space and their weighted averaging in the Fourier-space. The arithmetic average is used as this is the most simple and intuitive method, while the Fourier-based method is one of the best established in the field of multi-view microscopy[30,33,41] that preserves the most of the information content. The combined images are considerably shortened along the optical axis for the two fusion methods when compared to the originally elongated bead images, as illustrated by the intensity projection images (Fig. 4a-d); at the same time, the two fusion methods yield slightly different results. The enlarged regions of the 4 selected beads (Fig. 4e-h) emphasize more this radical shortening. They also illustrate the difference between the two fusion methods: the arithmetically averaged images are somewhat broader than the Fourier-averaged ones; as a result, images of two adjacent beads can be separated with higher contrast using the Fourier method (Fig. 4g inset). We have to note that in the Fourier-fused images small sidelobes tend to appear which are only about 5-15% of the bead image maxima. Considering that the Fourier-based method yields larger resolution improvement along the optical axis, it keeps all the image information and it is in accordance with the literature, we use this fusion method to prove the resolution improvement for the bead and for the mitochondrion-stained images.

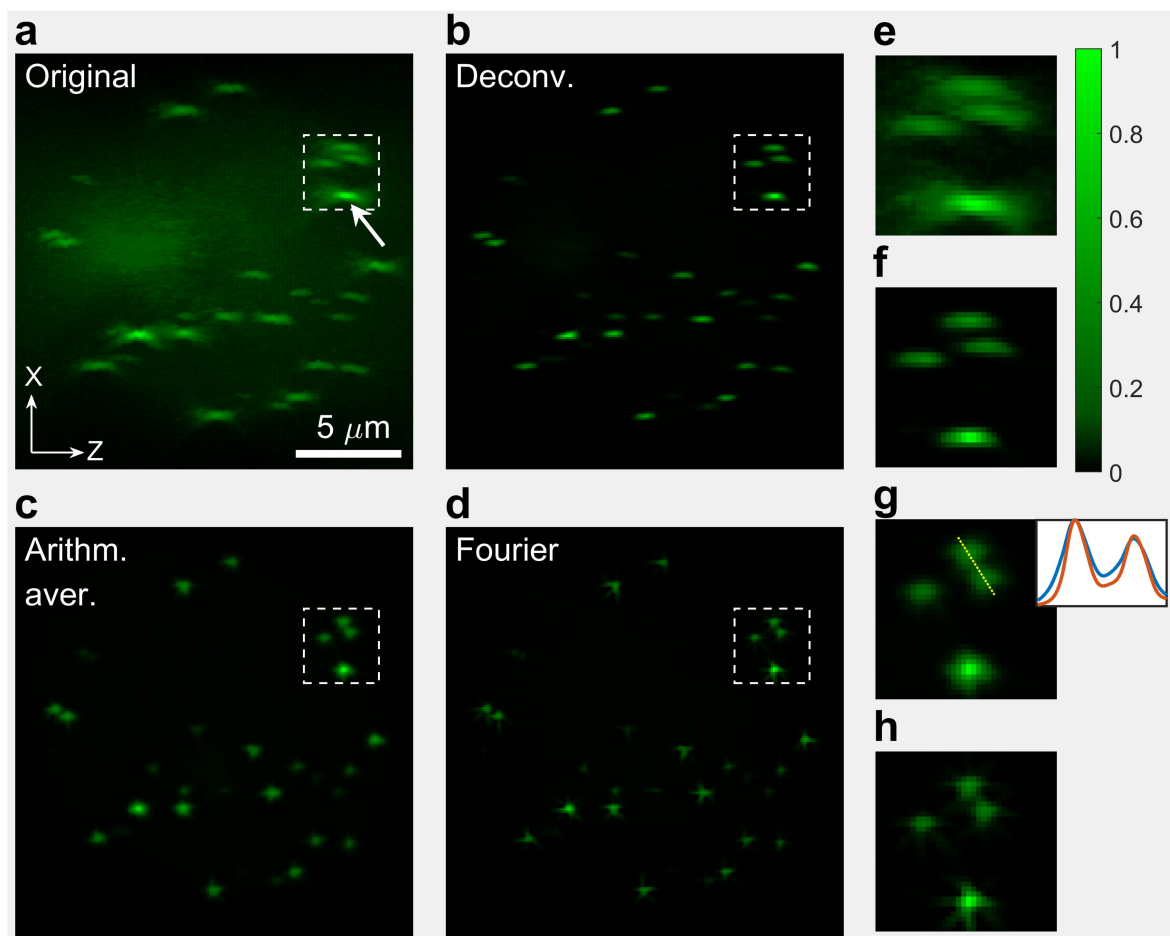

**Figure 4. Maximum intensity projections (MIP) of the 3D fluorescence intensity data arrays recorded on a trapped cell that was labelled with 100 nm fluorescent beads.** MIP of an original, unprocessed data array recorded on the entire cell (**a**), of a deconvolved array (26 iterations) (**b**), of a fused array with arithmetic averaging (0°, 45°, 90° and 135° orientations) (**c**) and of one fused with the Fourier transform-based method (**d**). **e-h** Enlarged MIP images of four beads in the highlighted areas of the original, deconvolved, averaged and Fourier-fused arrays, respectively. The curves on the inset of **g** show the intensity traces over two adjacent bead images from the arithmetically averaged (blue) and Fourier-fused (red) arrays along the line shown on **g**.

The resolution improvement, already evidenced from the maximum projection images was quantified by comparing the widths of selected bead images obtained again by Gaussian fitting along the three spatial axes (Fig. 5b) on the original, the deconvolved and the Fourier-fused arrays (the coordinate axes are those of the reference data array). The averaged 1/e width along the optical axis was reduced from 1.37 μm (original array) to 0.58 μm (fused array). This 2.4 times reduction is partly attributed to the deconvolution step but also stems from the fusion of the images taken at different orientations. The axial improvement was accompanied with a reduction in the lateral direction, which, however is only due to the deconvolution step. Interestingly, the fusion reversed this decrease by about 20%, which may be attributed to the mentioned fluctuation of the cell, to the not completely rigid attachment of some beads to the cell membrane or even to the slight imperfection of the alignment. The intensity traces shown on Fig. 5a also display convincingly that along the optical axis the width decreases during the whole process while in the lateral direction mostly during the deconvolution step.

## *3D structure of labelled single cell*

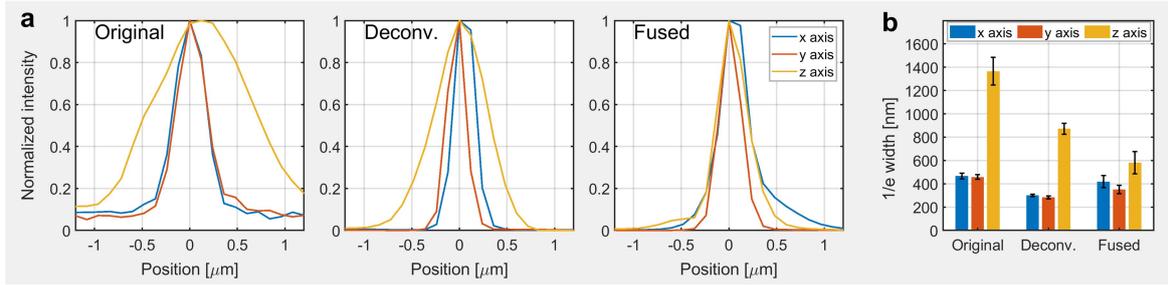

**Figure 5. The resolution improvement characterized with Gaussian fit of line traces along bead images. a** Normalized intensity traces over the bead image marked on Fig. 4a with the arrow, along the three axes taken from the original, deconvolved and fused arrays. **b** Average 1/e widths calculated with selected bead images from the original, deconvolved and Fourier-fused arrays along the three axes. The widths were determined with a Gaussian fit; the fusion included 4 orientations.

For the mitochondrion-stained arrays, the deconvolution was stopped after 60 iterations because the power spectrum of this array approximates properly the modulation transfer function (MTF) of the system PSF. We note here that the number of iterations in the literature is very often chosen according to subjective criteria, although its choice affects the contrast of the resulted images and therefore the feature sizes of the fused array. The real relevance of resolution improvement lies in the imaging of biological objects. Therefore, we also demonstrated it with imaging cell organelles: we selectively stained the cell's mitochondria (Fig. 6) and reconstructed it in 3D. Based on our previous findings, we fused arrays from only 4 orientations (0, 45, 90 and 135 degrees) to reconstruct the mitochondria's 3D arrangement. Fig. 6a visualizes the steps of the resolution enhancing procedure. From the selected 2D slices it is obvious that the deconvolution (Fig. 6a, second column) removes most of the out-of-focus noise and that the Fourier-based fusion separates merged spots corresponding to the roughly 0.5 μm large mitochondria along the Z axis (Fig. 6a, third column). Resolution improvement is also evidenced from the line traces (Fig. 6b) which could resolve those features that are not separable otherwise. The traces along the x axis on Fig. 6b (lowest graph) are only slightly different after deconvolution and fusion, indicating that fusion leaves the image patterns mostly unchanged in the lateral direction. While the resolution improvement in case of bead-decorated samples can be directly determined from measurements in the image space, this is not possible in the case of the mitochondrion-stained sample. Therefore, the resolution here was evaluated by calculating the spatial frequency power spectra of the arrays (Fig. 6c). The spectrum of a single deconvolved array shows the typical asymmetric shape of the MTF of the imaging system, being much narrower in the $k_z$ direction. However, the spectrum of the fused array (4 directions) shows a strong radial symmetry with a significant broadening in the $k_z$ direction, indicating more isotropic resolution.

### 3. Discussion

In the presented work, we combined multiview wide-field fluorescent observation and microstructure-assisted optical micromanipulation to realize resolution improved imaging of single cells. The stable cell-to-structure attachment and the high structure-to-water refractive index contr ast ensured the precise actuation of the cell and its high spatial stability throughout the imaging process. The use of the two-photon polymerized, functionalized 3D microstructures of task-specific shape guarantees the fast and easy access of the various orientations for imaging. This imaging scheme can be implemented in any type of sample holder wherever optical trapping can be carried out without the physical movement of the holder itself or inserting any mechanical tool from outside. The image sequences can be successfully combined into a single 3D image array, which displays practically isotropic resolution, as proved with 100 nm fluorescent beads. The slightly lower resolution along the original Z axis is mainly attributed to cell or bead fluctuations (being smaller than the optical resolution). The trapping force that primarily determines the fluctuation can be increased trivially with higher trapping laser power, or with re-designed manipulator structure with smaller trapping spheres; using two manipulator structures that hold the cell from two opposite directions may also reduce fluctuation. Interestingly, with 45 degrees rotations, only 4 directions were enough to obtain the highest resolution improvement, the addition of further directions in between did not result in further increase. It is very important to emphasize that the optical manipulation that allows the access of these viewing orientations, is not limited to a single rotation axis, like the mechanical techniques where the entire sample is rotated but cells can be viewed from any direction. One should, however avoid orientations where the manipulator or some of its part blocks the view of the cell.

The method applied on cells with labelled mitochondrion also yielded resolution-enhanced images where some features could be resolved along the optical axis only due to the multiview recording. Here, the experiments were performed on fixed cells, which do not change their structure during the course of the data acquisition (approx. 1 min/orientation). In order to follow dynamic events with our method inside live cells the data acquisition must be accelerated and cell fixation avoided. Faster acquisition would require selected plane fluorescent excitation where the light sheet is swiftly scanned across the cell for sectioning and the optical manipulation is used only to reach the various orientations for multiview recording. This scheme, with the extra benefit of reduced phototoxicity, could be combined with remote refocusing, and could lead to multi-orientation acquisition time of less than a minute. When live cells are used, their attachment to the manipulators must be done with extra care: cell treatment should be avoided and the biochemical agents used for the functionalization of the manipulators should not affect the physiology of the cells.

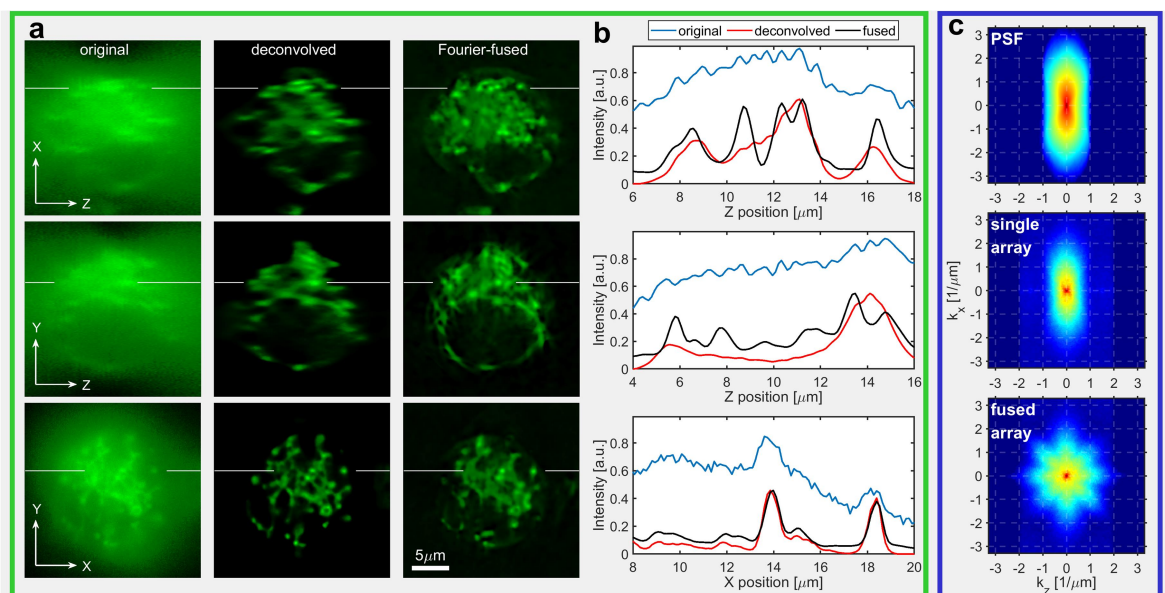

**Figure 6. Resolution improvement measured on mitochondria-stained single cells. a** Corresponding single slices taken from the original, deconvolved and fused 3D data arrays (columns) along the three axes (rows). **b** Intensity traces between the thin white lines on the single image slices of the corresponding rows. **c** MTF calculated from the measured point spread function (PSF), and power spectra obtained from a single array (deconvolved, 60 iterations) of a stained cell and from a fused array obtained with the Fourier-transform based method using four deconvolved arrays ($0°$, $45°$, $90°$ and $135°$ orientations).

Finally, we emphasize that the procedure introduced here can be implemented in any other microscopies where imaging is performed with a single objective; most notably, comparable resolution improvement can be achieved also in light sheet or confocal microscopy. Optical trapping requires the use of high NA (1.2) objective; the fact, that this objective is also used for imaging, similarly to the most up-to-date multiview microscopy setups[32,42], ensures the presented high resolution. In comparison, multiview systems generally use low or medium NA, sometimes water dipping, objectives for imaging (from 0.2 up to 1) due to the required large working distance[43]. A potentially important future application of the optical manipulation method is the observation of cell-cell interactions with isotropic resolution, with the added benefit of the accurate timing of the reaction and its observation. The possibility of rotation of the cell around any axis can ensure unprecedented spatial control of the mutual position of the two cells where the contact surface can be freely selected.

4. Methods

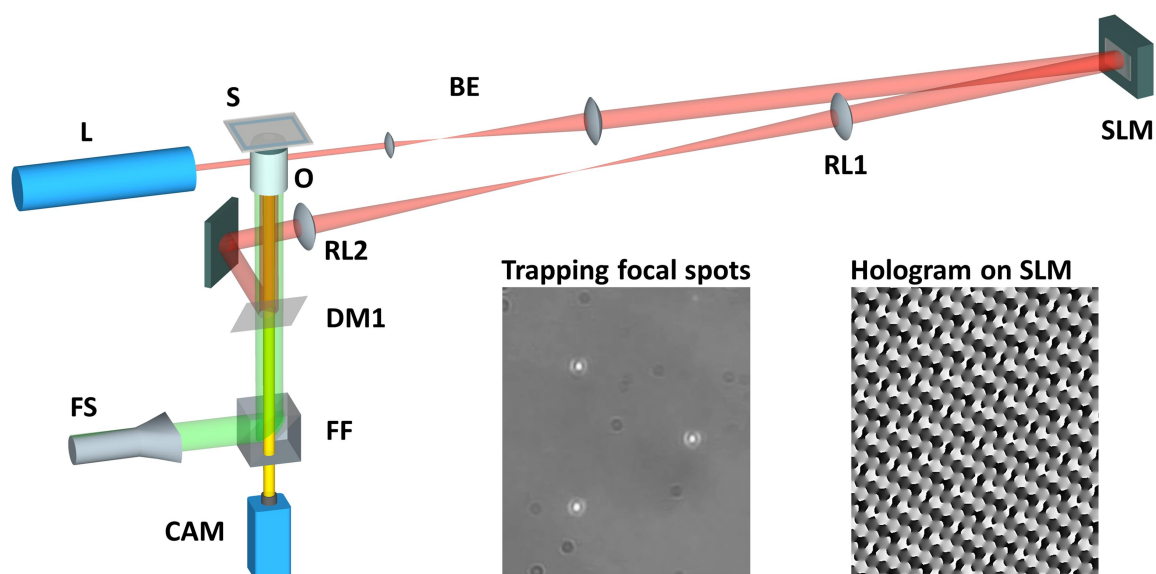

**Figure 7. The optical layout of the imaging system.** L: trapping laser source, BE: beam expander to slightly overfill the SLM surface, SLM: spatial light modulator, RL1 and RL2: relay lenses to project the SLM surface to the entrance pupil of the objective, DM: dichroic mirror that reflects the trapping beam into the objective but transmits the fluorescence beams, FF: fluorescence filter set, FS: fluorescent light source, CAM: camera, O: objective, S: sample. The inserted hologram creates the three trapping focal spots in the sample plane.

The experimental setup consists of the combination of a conventional, wide-field fluorescent microscope completed with holographic optical tweezers, as displayed in Fig. 7. Here, the same high NA microscope objective is used to trap and manipulate the cell-attached microtools, as well as to excite the fluorophores and to collect the emitted light. We used the optical tweezers to rotate the cells into pre-defined orientations, to translate them through the observation plane and to hold them steady during the fluorescent imaging. The access of these positions by the indirect trapping allowed for the imaging of the cells with isotropic resolution in their aqueous environment.

*Microtool design*

The shape of the microtools was directed by the experimental details with regard to optimal manipulation efficiency. The structure is held by three optical traps produced by the holographic optical tweezers, the points of trapping are provided by the three spheres of the structure with diameter of 4 µm (Fig. 8, insert). The number of the trapping spheres was reduced to the minimal three required for 6 degrees of freedom (6DoF) actuation. Still, very good stability of the cell position could be achieved due to the following conditions: The cell is held about 10 micrometers away from the trapping light beams such a way that the symmetry axis of the microtool-cell complex is perpendicular to the optical path (Fig. 1b and Fig. 8). Consequently, both lateral and axial motions of the cell translate into primarily lateral motions of the trapped microtool spheres; the trapping forces are known to be the highest for displacement in this direction ensuring maximal stability. Furthermore, the cell is attached to the structure through a concave disk of 8 µm radius of curvature, designed to accommodate the spherical shape of the cells of 12-20 µm diameter for maximum attachment probability and stability.

The structures were built by two-photon polymerization (TPP). TPP was carried out in the system described in detail recently[44]. The SU-8 2007 material (Microchem GmbH, Germany) was polymerized with the light from an ultrashort-pulsed laser ($\lambda$ = 785 nm, pulse length = 100 fs, repetition rate = 100 MHz, C-Fiber A, Menlo Systems, Germany). The single beam of the laser was multiplied into four beams with a spatial light modulator (Pluto NIR, Holoeye GmbH, Germany) and these steady beams were focused by a Zeiss Achroplan 100X 1.25 NA oil immersion objective into a thin layer of the resin atop a microscope cover slide that was translated in 3D by a piezo scanner system (P-124 731.8L and P-721.10, Physik Instrumente GmbH, Germany). The scanning speed was between 1 and 32 µm/s at a laser intensity of 3 mW. Standard procedure was applied for the development of the illuminated resin: post-bake at 95 °C for 10 mins, development in mrDev 600 (Microchem GmbH, Germany), followed by rinsing in ethanol. The insert in Fig. 8 shows the scanning electron microscopic image of a prepared structure. The parallel polymerization scheme enabled us to routinely fabricate about 1500 structures overnight, generally enough for a dozen of samples. The application of similar two-photon polymerized trapped structures has been published already by others[19,45] and by us[20,46,47].

*Microstructure functionalization, cell attachment and labeling*

Attachment of the cells to the structures was achieved by the affinity of the protein concanavalin A (ConA) towards the glycocalix of mammalian cells. The surface of the structures had to be functionalized but in contrast to our earlier work[20] the cells themselves didn't need to be treated. Shortly, the functionalization of the SU-8 microtools started by an acid-treatment (30 mins incubation in the mixture of 1M nitric acid and 0.1M cerium(IV)-ammonium nitrate on RT) and a subsequent PEG-diamine coating (15 mM methanol solution of MW 2000 PEG-diamine on RT). In the next steps sulpho-NHS biotin (90 mins incubation in 1 mg/mL PBS solution on RT) and streptavidin (overnight in 100 nM PBS solution at 4 °C) were bound to the surface. Finally, the structures were incubated in 1 mg/mL solution of biotinated ConA for 1 hr at 4 °C, then washed thoroughly with water and dried.

We used fixed K562 white blood cells in our experiments (LGC Standards, UK, cat no. CCL-243); the fixation was carried out as follows. First, the cell suspension was centrifuged twice for 7 mins at 125 RCF each to replace their growth medium with PBS; they were resuspended PBS to obtain ~$10^6$ cell/mL. 750 µL of this suspension was mixed with 750 µL of 4% formaldehyde in PBS and incubated at 4 °C for 20 mins while shaking. It was followed by two washing steps with centrifuge parameters as before. Eventually the cells were resuspended for the desired density (~$10^6$ cell/mL) with PBS. The fixed cells were then decorated with fluorescent beads or stained with an organelle-specific fluorophore.

Measurements with bead-based resolution estimations were performed with 100 nm diameter, cell-attached fluorescent beads (CAFR100NM, Magsphere Inc., USA, exc. max.: 540 nm, em. max.: 584 nm), the diameter of which was below the resolution of the imaging system. First, 1 µL of the original bead solution was diluted 500x in PBS buffer, then centrifuged (14000 RCF) for 30 mins followed by the careful removal of 90% of the supernatant obtaining ~50 µL suspension; this second step was repeated twice. Next, 1 mL of the fixed cells was incubated with the entire bead suspension for 45 mins in dark. After the incubation, the cells were again washed in PBS twice for 3 mins each to remove the unbound beads. For the demonstration of isotropic cell organelle imaging with our method, the cells' mitochondria were fluorescently stained (Mitotracker Deep Red, Thermo Fisher Scientific, exc.: 642 nm, em.: 662 nm). First, the mitochondrion stain was added into 1 mL of cell suspension to reach 6000x dilution, then the cells were incubated with it for 1 hr in dark and finally a washing step was applied similarly to the last step of bead attachment. Once the cells were labelled, they were collected in PBS containing 0.1 % Tween20 surfactant to prevent their adhesion to the glass walls of the sample holder. Approximately 10 µL of this suspension was dropped onto the functionalized micromanipulators, the droplet was gently stirred to remove the structures from their glass substrate and about 5 µL of this cell-manipulator mixture was pipetted onto a cleaned coverslide (24 mm x 40 mm). A two-sided tape was used as spacer and a second coverslide was used to close the sample chamber; the cell manipulation then took place within an approximately 80 µm high liquid layer of the sandwiched sample.

*Microscopy and optical trapping*

The optical trapping-assisted fluorescence observation was performed in an extended Nikon Eclipse TI inverted fluorescence microscope. The complete optical layout including the trapping beam management and the fluorescent excitation/observation path is shown on Fig. 7. Light for the optical tweezers came from a continuous wave fiber laser (L, λ=1070 nm, P=10W, THFL-1P400-COL50, BKtel Photonics, France). The holographic optical tweezers (HOT) was based on a reflective phase only spatial light modulator (SLM, PLUTO NIR, Holoeye, Germany). A 4f lens system (lenses RL1 and RL2) was used to project the plane of the SLM to the back aperture of the objective[48] and a high NA microscope objective (O on Fig. 7, 60X, water immersion, NA=1.2, Olympus UPlanSApo) to generate the trapping foci in the sample (S). The near infrared light of the optical tweezers was coupled into the microscope with a low pass dichroic mirror (DM, cut off: λ=850 nm) so it did not interfere with the fluorescence excitation/observation. This optical system enabled the generation, and the totally controlled 3D motion of 3 trapping focal spots. The highest performance for the holographic optical traps was ensured by applying a wavefront correction hologram for the complete trapping optical system determined with the method of Cizmar[49].

For the required velocity of the manipulation of multiple foci, a fast calculation of the beam-shaping holograms was needed. We used an NVIDIA CUDA GPU (GeForce GTX 660 with 960 CUDA cores) to calculate the holograms as described in detail in[44], implementing the weighted Gerchberg-Saxton (GSW) algorithm[48]. With our procedure, we were able to calculate up to 10 foci with 10 iterations within the refresh cycle of our SLM (16.67 ms). This rate is sufficient for the real time continuous repositioning of the 3 traps necessary for the orientation of our microtools. We created a special user program for the actuation of the cell manipulators and for the synchronized control of the fluorescence recording camera. Eventually our system enabled the direct and real time manipulation of the cells with 6DoF. We note that manipulation of a single structure with 3 holographic traps can be achieved with the simple "prisms and lenses" algorithm[48] that requires only CPU-based calculation for real time speed. Also, a GPU-based implementation of the GSW algorithm is publicly available (open-source) from: https://github.com/MartinPersson/HOTlab.

The fluorescent part of the microscope used a metal halide light source for excitation (FS on Fig. 7, model: Lumen 200S, Prior Scientific, Inc., USA) and filter sets (FF) that match the fluorescence characteristics of the fluorescent beads or that of the fluorophore used for cell staining (filter sets 49008, exc.: 560/40 nm, em.: 630/70 nm for the beads and 49006, exc.: 620/50 nm, em.: 700/70 nm for the cellular stain, both from Chroma Technology Corp., USA). The fluorescent images were recorded by an 1004x1002 pixels EMCCD camera (CAM, Rolera EMC$^2$, Qimaging, Canada), controlled by our trapping software; the imaging train defines 120 nm lateral pixel distance on the recorded images. The bright-field images were taken with a CCD camera (GS3-U3-23S6M, Point Grey Research Inc., Canada).

*Trapping of a cell and data acquisition*

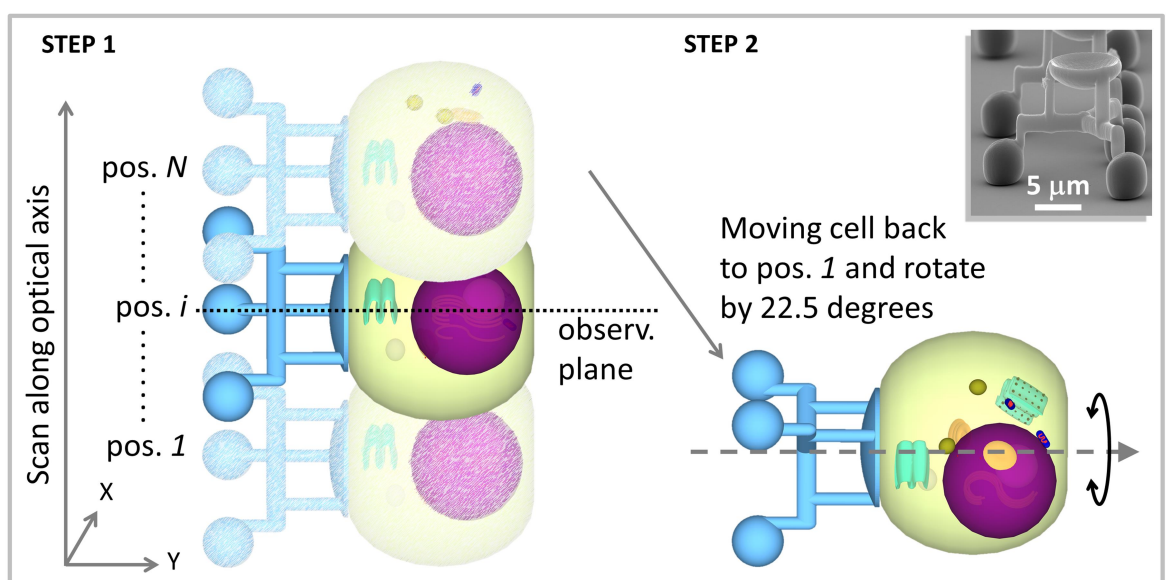

**Figure 8. The concept of the application of the cell manipulator microtool for multiview imaging.** Step 1: HOT-assisted data acquisition via alternating cell translations to pre-defined positions (pos. 1 through pos. N) and image recordings at each position. Step 2: return the cell to pos. 1 and its rotation to a new orientation. Inset: SEM image of cell manipulator microtools.

The data acquisition process included fluorescent sectioning of the cell along the optical axis at various orientations and its rotation to these orientations, repeatedly. It was the task of the HOT to move the cell along the z axis to pre-defined positions for sectioning, to hold the cell as steady as possible at these positions during image acquisition and finally to rotate it precisely to the orientations required for the multiple-angle imaging. Before these, however, a free-floating cell had to be attached to the functionalized and trapped microtool. The procedure, illustrated on Fig. 2a-c, was carried out similarly as reported in our earlier works[20,21]. First, the structure was trapped with the three trapping foci. Then its attachment disk was oriented towards the cell. Finally, it was pushed against the cell by its attachment disk for a few seconds while the ConA bound to the cells surface; the cell is now indirectly trapped and free to be actuated through the manipulator with 6DoF. 90 mW laser power was measured for each trapping beam at the objective entrance pupil; the transmittance of the objective at 1070 nm is approx. 50%, so each beam delivers ~45 mW in the focus. After taking hold of the cell, it was oriented such that the symmetry axis of the cell-microtool complex was perpendicular to the optical axis (Fig. 1b). The complex was rotated around this symmetry axis (similarly to Fig. 2c-d) during the acquisition process. The lateral positional stability of the indirectly trapped cell was determined by video-microscopic imaging of the cell-microtool complex and image analysis using a built-in Matlab (The Mathworks Inc, Natick, Ma, USA) local feature detector function "`detectSURFFeatures`".

The imaging process, as illustrated in Fig. 8 started with the fluorescent sectioning (z-scan) of the cell at the first orientational position (0 degree). Here, the cell was translated with the HOT along the optical axis through the static observation plane (microscope objective is not moving) in 250 nm steps (Step 1 on Fig. 8). At each vertical position, 3 fluorescent images were averaged with the EM-CCD camera using 100-120 ms integration time. The EM gain and integration time were set to avoid image saturation. In a typical sectioning images were recorded at about 80 positions spanning the approx. 20 µm diameter of the cell. After recording one stack, the cell was translated to its starting position and rotated 22.5 degrees with the optical trap (Step 2 on Fig. 8) for the next sectioning sequence. This process was repeated 8 times in the fluorescent bead experiments to complete a half rotation yielding altogether 9 sets of data (0-180degres). In the experiments imaging cells with the stained mitochondria the process was carried out at 5 rotational positions from 0 to 180 degrees with 45 degrees steps. The imaging in each rotational position resulted in a 3D data array of $m$-by-$n$-by-z dimension, where $m$ and $n$ are the pixel size of each frame and $z$ is the number of frames; $m$ and $n$ were typically 400.

*Data evaluation procedure*

The data evaluation process on the image stacks was adopted from the literature of multiview microscopy[30,33,50]. It consisted of three main steps: *i)* pre-processing, where the out of focus intensity signals were largely removed from the data arrays by deconvolution; *ii)* the registration of the image stacks obtained from the different directions; *iii)* fusion of the deconvolved images to yield the single output 3D image with improved resolution. All steps were performed with home-developed routines coded in Matlab; the calculations were performed on an NVIDIA Titan XP GPU-equipped desktop PC relying on the built-in CUDA support of the Matlab functions.

The 3D deconvolution of each data array was performed using a PSF measured with the fluorescence filter sets used in the experiments; nanoholes of ~100 nm diameter, fabricated into a reflective gold layer were used as point-like light sources. We used the iterative Richardson-Lucy (RL) deconvolution algorithm. The iteration numbers were determined for beads from the relative change of the consecutive deconvolved stacks and from the dimensions of the deconvolved bead images obtained with a Gaussian fitting and for the mitochondrion-stained cell from the power spectrum of the deconvolved array. The determined iteration number for the fluorescent bead experiment was 26 and for the mitochondria-stained sample 60.

The data was trimmed prior to the registration as follows: first, the 3D arrays were cropped leaving only the cells and its direct surrounding on the pictures; second, an intensity threshold was applied to keep only the about 5000 (fluorescent beads) or about the 500000 (mitochondrion stain) most intense pixels in order to speeds up our algorithm considerably by leaving out useless information; third, each data array was interpolated along their own z axes so that the data points are separated by 120 nm in this direction to be equal to the lateral pixel distance.

In the image registration procedure the deconvolved stacks of different orientations were precisely aligned using a correlation-based algorithm in Matlab. In the first step of registration the data arrays were manually rotated around the experimental rotation axis by the angles used to rotate the structures in the measurements (multiples of 22,5°) leaving the first measurement (0º) as reference. Then, for the precise registration rigid transformations, i.e. translation and rotation of the data arrays were applied with an iterative direct search method. Here, first rotation axes were defined in the 2π steradian solid angle by θ (polar) and φ (azimuth) angles, with 45º angular distance. The manually rotated array was then rotated around these few axes in the -15º to +15º range in 1º steps and the translational correlation between the rotated and the reference stacks was calculated for each position. Then, the axis resulting in the maximum correlation coefficient was chosen and a set of new axes were defined in the solid angle between this and its neighboring axes with the half of the angular distance as before. In the next iteration step, the manually rotated stack was rotated around these new axes and the correlation was calculated again. This iterative process always converged within 10 iterations, which was confirmed by the maximum correlation coefficient showing a plateau. The process eventually yielded an axis of rotation, a rotation angle around it and 3 translation values along the Descartes axes that showed the highest correlation between the reference and the aligned array.

Finally, the images were fused to yield a single 3D stack as an output. Here, the deconvolved and interpolated, but not cropped or thresholded data arrays of the different directions were first transformed by the determined translations and rotations. For some samples we observed photobleaching of the cells, which was corrected for before fusion with a normalization step: the data arrays were normalized to the reference one with the integrated fluorescent intensity of their 1500 most intense pixels. Two fusion methods were compared to find the final intensity of any given pixel location: the arithmetic average of the pixel intensities of the aligned arrays, and the weighted average of the 3D Fourier transforms of the stacks were calculated at each location of the frequency space using the following equation[50]:

$$\tilde{I} = \sum_{j=1}^{N} \left( w_j \cdot \tilde{I}_j \right) \quad \text{Eq. 2.}$$

where $N$ is the number of stacks, $\tilde{I}_j$ is the spectral data and $w_j$ are the weights calculated as follows:

$$w_j = \frac{\sqrt{|\tilde{I}_j|}}{\sum_{j=1}^{N}\left(\sqrt{|\tilde{I}_j|}\right)}. \quad \text{Eq. 3.}$$

Finally, the $\tilde{I}$ average was inverse Fourier-transformed to get the reconstructed 3D array in the real space.

**References**

1    Abbondanzieri, E. A., Greenleaf, W. J., Shaevitz, J. W., Landick, R. & Block, S. M. Direct observation of base-pair stepping by RNA polymerase. *Nature* **438**, 460-465, doi:10.1038/nature04268 (2005).
2    Meiners, J. C. & Quake, S. R. Femtonewton force spectroscopy of single extended DNA molecules. *Phys Rev Lett* **84**, 5014-5017, doi:10.1103/PhysRevLett.84.5014 (2000).
3    Brouwer, I. *et al.* Sliding sleeves of XRCC4–XLF bridge DNA and connect fragments of broken DNA. *Nature* **535**, 566, doi:10.1038/nature18643 (2016).

4       Smith, S. B., Cui, Y. J. & Bustamante, C. Overstretching B-DNA: The elastic response of individual double-stranded and single-stranded DNA molecules. *Science* **271**, 795-799, doi:DOI 10.1126/science.271.5250.795 (1996).
5       Cecconi, C., Shank, E. A., Dahlquist, F. W., Marqusee, S. & Bustamante, C. Protein-DNA chimeras for single molecule mechanical folding studies with the optical tweezers. *Eur Biophys J Biophy* **37**, 729-738, doi:10.1007/s00249-007-0247-y (2008).
6       Kellermayer, M. S. Z., Smith, S. B., Granzier, H. L. & Bustamante, C. Folding-unfolding transitions in single titin molecules characterized with laser tweezers. *Science* **276**, 1112-1116, doi:10.1126/science.276.5315.1112 (1997).
7       Finer, J. T., Simmons, R. M. & Spudich, J. A. Single Myosin Molecule Mechanics - Piconewton Forces and Nanometer Steps. *Nature* **368**, 113-119, doi:10.1038/368113a0 (1994).
8       Svoboda, K., Schmidt, C. F., Schnapp, B. J. & Block, S. M. Direct Observation of Kinesin Stepping by Optical Trapping Interferometry. *Nature* **365**, 721-727, doi:10.1038/365721a0 (1993).
9       Ashkin, A., Dziedzic, J. M. & Yamane, T. Optical Trapping and Manipulation of Single Cells Using Infrared-Laser Beams. *Nature* **330**, 769-771, doi:10.1038/330769a0 (1987).
10      Zhang, H. & Liu, K. K. Optical tweezers for single cells. *J R Soc Interface* **5**, 671-690, doi:10.1098/rsif.2008.0052 (2008).
11      McAlinden, N., Glass, D. G., Millington, O. R. & Wright, A. J. Accurate position tracking of optically trapped live cells. *Biomed Opt Express* **5**, 1026-1037, doi:10.1364/Boe.5.001026 (2014).
12      Lopez-Quesada, C. *et al.* Artificially-induced organelles are optimal targets for optical trapping experiments in living cells. *Biomed Opt Express* **5**, 1993-2008, doi:10.1364/Boe.5.001993 (2014).
13      Ayano, S., Wakamoto, Y., Yamashita, S. & Yasuda, K. Quantitative measurement of damage caused by 1064-nm wavelength optical trapping of Escherichia coli cells using on-chip single cell cultivation system. *Biochem Bioph Res Co* **350**, 678-684, doi:10.1016/j.bbrc.2006.09.115 (2006).
14      Liang, H. *et al.* Wavelength dependence of cell cloning efficiency after optical trapping. *Biophys J* **70**, 1529-1533, doi:10.1016/S0006-3495(96)79716-3 (1996).
15      Liu, Y. *et al.* Evidence for Localized Cell Heating Induced by Infrared Optical Tweezers. *Biophys J* **68**, 2137-2144, doi:10.1016/S0006-3495(95)80396-6 (1995).
16      Neuman, K. C., Chadd, E. H., Liou, G. F., Bergman, K. & Block, S. M. Characterization of photodamage to Escherichia coli in optical traps. *Biophys J* **77**, 2856-2863, doi:10.1016/S0006-3495(99)77117-1 (1999).
17      Rasmussen, M. B., Oddershede, L. B. & Siegumfeldt, H. Optical tweezers cause physiological damage to Escherichia coli and Listeria bacteria. *Appl Environ Microb* **74**, 2441-2446, doi:10.1128/Aem.02265-07 (2008).
18      Turlier, H. *et al.* Equilibrium physics breakdown reveals the active nature of red blood cell flickering. *Nat Phys* **12**, 513, doi:10.1038/nphys3621 (2016).
19      Phillips, D. B. *et al.* Force sensing with a shaped dielectric micro-tool. *Epl-Europhys Lett* **99**, 58004, doi:10.1209/0295-5075/99/58004 (2012).
20      Aekbote, B. L. *et al.* Surface-modified complex SU-8 microstructures for indirect optical manipulation of single cells. *Biomed Opt Express* **7**, 45-56, doi:10.1364/Boe.7.000045 (2016).
21      Vizsnyiczai, G. *et al.* in *Optical Trapping and Optical Micromanipulation Xiii* Vol. 9922 *Proceedings of SPIE* (eds K. Dholakia & G. C. Spalding)  992216 (2016).

22   Hell, S. & Stelzer, E. H. K. Fundamental improvement of resolution with a 4Pi-confocal fluorescence microscope using two-photon excitation. *Opt Commun* **93**, 277-282, doi:10.1016/0030-4018(92)90185-T (1992).
23   Choi, H. *et al.* Improvement of axial resolution and contrast in temporally focused widefield two-photon microscopy with structured light illumination. *Biomedical Optics Express* **4**, 995-1005, doi:10.1364/BOE.4.000995 (2013).
24   Sheppard, C. J. R. & Gu, M. Improvement of axial resolution in confocal microscopy using an annular pupil. *Opt Commun* **84**, 7-13, doi:10.1016/0030-4018(91)90019-A (1991).
25   Siegel, N. & Brooker, G. Improved axial resolution of FINCH fluorescence microscopy when combined with spinning disk confocal microscopy. *Optics Express* **22**, 22298-22307, doi:10.1364/OE.22.022298 (2014).
26   Fouquet, C. *et al.* Improving Axial Resolution in Confocal Microscopy with New High Refractive Index Mounting Media. *Plos One* **10**, doi:10.1371/journal.pone.0121096 (2015).
27   Heintzmann, R. & Cremer, C. Axial tomographic confocal fluorescence microscopy. *Journal of Microscopy-Oxford* **206**, 7-23, doi:10.1046/j.1365-2818.2002.01000.x (2002).
28   Chen, Y. *et al.* Effects of axial resolution improvement on optical coherence tomography (OCT) imaging of gastrointestinal tissues. *Opt Express* **16**, 2469-2485, doi:10.1364/OE.16.002469 (2008).
29   Wolfson, D. *et al.* Rapid 3D fluorescence imaging of individual optically trapped living immune cells. *Journal of Biophotonics* **8**, 208-216, doi:10.1002/jbio.201300153 (2015).
30   Shaw, P. J., Agard, D. A., Hiraoka, Y. & Sedat, J. W. Tilted View Reconstruction in Optical Microscopy - 3-Dimensional Reconstruction of Drosophila-Melanogaster Embryo Nuclei. *Biophys J* **55**, 101-110, doi:10.1016/S0006-3495(89)82783-3 (1989).
31   Huisken, J., Swoger, J., Del Bene, F., Wittbrodt, J. & Stelzer, E. H. K. Optical Sectioning Deep Inside Live Embryos by Selective Plane Illumination Microscopy. *Science* **305**, 1007-1009, doi:10.1126/science.1100035 (2004).
32   Krzic, U., Gunther, S., Saunders, T. E., Streichan, S. J. & Hufnagel, L. Multiview light-sheet microscope for rapid in toto imaging. *Nat Methods* **9**, 730, doi:10.1038/nmeth.2064 (2012).
33   Swoger, J., Huisken, J. & Stelzer, E. H. Multiple imaging axis microscopy improves resolution for thick-sample applications. *Opt Lett* **28**, 1654-1656, doi:10.1364/OL.28.001654 (2003).
34   Friese, M., Nieminen, T., Heckenberg, N. & Rubinsztein-Dunlop, H. Optical alignment and spinning of laser-trapped microscopic particles (vol 394, pg 348, 1998). *Nature* **395**, 621-621, doi:10.1038/27014 (1998).
35   La Porta, A. & Wang, M. Optical torque wrench: Angular trapping, rotation, and torque detection of quartz microparticles. *Physical Review Letters* **92**, doi:10.1103/PhysRevLett.92.190801 (2004).
36   Kolb, T., Albert, S., Haug, M. & Whyte, G. Dynamically reconfigurable fibre optical spanner. *Lab on a Chip* **14**, 1186-1190, doi:10.1039/c3lc51277k (2014).
37   Forrester, A., Courtial, J. & Padgett, M. Performance of a rotating aperture for spinning and orienting objects in optical tweezers. *Journal of Modern Optics* **50**, 1533-1538, doi:10.1080/0950034021000020798 (2003).
38   Kelemen, L., Valkai, S. & Ormos, P. Integrated optical motor. *Applied Optics* **45**, 2777-2780, doi:10.1364/AO.45.002777 (2006).

39  Galajda, P. & Ormos, P. Complex micromachines produced and driven by light. *Applied Physics Letters* **78**, 249-251, doi:10.1063/1.1339258 (2001).
40  Kolb, T., Albert, S., Haug, M. & Whyte, G. Optofluidic rotation of living cells for single-cell tomography. *Journal of Biophotonics* **8**, 239-246, doi:10.1002/jbio.201300196 (2015).
41  Fei, P. *et al.* Subvoxel light-sheet microscopy for high-resolution high-throughput volumetric imaging of large biomedical specimens. *Advanced Photonics* **1**, 1-13, 13 (2019).
42  Kolb, T., Albert, S., Haug, M. & Whyte, G. Optofluidic rotation of living cells for single-cell tomography. *J Biophotonics* **8**, 239-246, doi:10.1002/jbio.201300196 (2015).
43  Greger, K., Swoger, J. & Stelzer, E. H. K. Basic building units and properties of a fluorescence single plane illumination microscope. *Rev Sci Instrum* **78**, doi:10.1063/1.2428277 (2007).
44  Vizsnyiczai, G., Kelemen, L. & Ormos, P. Holographic multi-focus 3D two-photon polymerization with real-time calculated holograms. *Opt Express* **22**, 24217-24223, doi:10.1364/Oe.22.024217 (2014).
45  Simons, M. *et al.* Directly interrogating single quantum dot labelled UvrA2 molecules on DNA tightropes using an optically trapped nanoprobe. *Sci Rep-Uk* **5**, 18486, doi:10.1038/srep18486 (2015).
46  Palima, D. *et al.* Wave-guided optical waveguides. *Opt Express* **20**, 2004-2014, doi:10.1364/Oe.20.002004 (2012).
47  Vizsnyiczai, G. *et al.* Optically Trapped Surface-Enhanced Raman Probes Prepared by Silver Photoreduction to 3D Microstructures. *Langmuir* **31**, 10087-10093, doi:10.1021/acs.langmuir.5b01210 (2015).
48  Di Leonardo, R., Ianni, F. & Ruocco, G. Computer generation of optimal holograms for optical trap arrays. *Opt Express* **15**, 1913-1922, doi:10.1364/Oe.15.001913 (2007).
49  Cizmar, T., Mazilu, M. & Dholakia, K. In situ wavefront correction and its application to micromanipulation. *Nat Photonics* **4**, 388-394, doi:10.1038/Nphoton.2010.85 (2010).
50  Swoger, J., Verveer, P., Greger, K., Huisken, J. & Stelzer, E. H. K. Multi-view image fusion improves resolution in three-dimensional microscopy. *Opt Express* **15**, 8029-8042, doi:10.1364/Oe.15.008029 (2007).

**Acknowledgements**

We are grateful for Péter Galajda (Institute of Biophysics, Biological Research Centre) for making his microscope system available for the extension with HOT. This work was supported by the GINOP-2.3.2-15-2016-00001 and the GINOP-2.3.3-15-2016-00040 programs. This project also received funding from the European Union's Horizon 2020 research and innovation programme under grant agreement No. 654148 Laserlab-Europe. We gratefully acknowledge the support of NVIDIA Corporation with the donation of the Titan Xp GPU used for this research.

**Conflict of interest**

The authors declare that they have no conflict of interest.

**Contributions**

G.V., and L.K. conceived the concept and designed the experiments. T.F., I.G., and B.L.A, fabricated and functionalized the microstructures and performed the trapping experiments. T.F., and I.G. maintained the cell culture. G.V. designed and built the optical trapping setup. G.V, A.B., and L.K. performed the data evaluation process. G.V., P.O., and L.K wrote the paper and all the authors read and edited the paper.